\documentstyle[twoside,fleqn,espcrc2,fps]{article}

\newcommand{\AmS}{{\protect\the\textfont2
  A\kern-.1667em\lower.5ex\hbox{M}\kern-.125emS}}

\title{Improved Overlap Fermions}

\author{W. Bietenholz\address{NORDITA,
Blegdamsvej 17, DK-2100 Copenhagen \O, Denmark}
        and 
I. Hip\address{Institut f\"{u}r Theoretische Physik, Universit\"{a}t Graz,
A-8010 Graz, Austria}
\thanks{Talk presented by W.B. at LATTICE99.}
}
\begin{document}

\begin{abstract}

We test exact and approximate Ginsparg-Wilson fermions
with respect to their chiral and scaling behavior in the
2-flavor Schwinger model. We first consider explicit
approximate GW fermions in a short range, then we proceed
to their chiral correction by means of the ``overlap
formula'', and finally we discuss a numerically efficient
perturbative chiral correction. In this way we combine
very good chiral and scaling properties with a relatively
modest computational effort.

\vspace*{-7mm}

\end{abstract}

\maketitle

Recent work revealed that the Ginsparg-Wilson relation (GWR) \cite{GW}
\vspace*{-1mm}
\begin{displaymath}
\{ D_{x,y}, \gamma_{5} \} = 2 \, (D \gamma_{5} R D)_{x,y}
\end{displaymath}
\vspace*{-1mm}
provides the correct chiral behavior of the lattice fermion
characterized by the lattice Dirac operator $D$, if $R$ is
a {\em local} Dirac scalar \cite{ML}.
However, this relation does not imply anything about the
scaling quality. Our goal is the combination
of excellent chiral {\em and} scaling behavior -- as well
as practical applicability. 
For details, see Ref.\ \cite{BH}.

In perfect and classically perfect actions, the chiral symmetry
is represented correctly, since it is preserved under block variable
renormalization group transformations, so it can
be traced back to the continuum \cite{BW}.
This is in agreement with the fact that such actions solve the GWR.
For the classically perfect actions, this has been shown
in Ref.\ \cite{HLN}, and it is also known that they
scale excellently.
However, they cannot be applied immediately
since they involve an infinite
number of couplings. Unfortunately this seems to be true
for any solution of the GWR \cite{WB,ultraloc}.
A truncation, which is applicable to QCD \cite{HF}, is the 
``hypercube fermion'' (HF) with couplings to all sites in a unit hypercube.
We test the quality of suitable 2d HFs regarding:
\begin{itemize}
\vspace*{-2mm}
\item chirality: we focus on the form $2 R_{x,y} = \mu \ \delta_{x,y}$
($ \mu > 0$),
where the spectrum $\sigma (D)$ of an exact GW fermion lies on the
circle in $C \!\!\!\! \tiny{I}$ with radius and center $1/\mu$
(GW circle). We check how well this is approximated.
\vspace*{-2mm}
\item scaling: we test the fermionic and ``mesonic'' dispersion
relation.
\end{itemize}
\vspace*{-2.5mm}
Our general ansatz reads $D= \rho_{\nu}\gamma_{\nu} + \lambda$,
where the vector term $\rho_{\nu}$ is odd in $\nu$ direction and
even in the other direction, while the scalar term $\lambda$ is entirely even.
In the free case, $\rho_{\nu}(x-y)$ $\{ \lambda (x-y) \}$ contains 2 $\{ 3\}$ 
different couplings. We consider a massless ``scaling optimal'' HF (SO-HF), 
which is constructed by hand. Its free couplings are
$\rho_{1}(10) = 0.334$, $\rho_{1}(11) = 0.083$;
$\lambda (00) = 3/2$, 
$\lambda (10) = 2 \lambda (11) = -1/4$
(similar to a truncated perfect free fermion).
The free spectrum approximates well the GW circle for $\mu =1$ 
(standard GWR) \cite{BH}.
From the free fermion dispersion relation shown in Fig.\ 1, 
we see that the SO-HF is strongly improved over the Wilson fermion.
\begin{figure}[hbt]
\vspace{-9mm}
\def\fpsangle{270}
\epsfxsize=40mm
\fpsbox{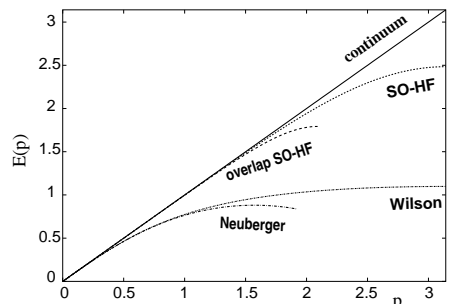}
\vspace{-11mm}
\caption{\it{Free fermion dispersion relations.}}
\vspace{-10mm}
\end{figure}

We now proceed to the 2-flavor Schwinger model, and we attach the free couplings
to the shortest lattice paths only (where there exist 2 shortest
paths, each one picks up half of the coupling). Moreover, we add a clover
term with coefficient 1, since this turned out to be useful.
\footnote{This is not renormalized in the Schwinger model \cite{HLT}.}
Typical configurations on a $16 \times 16$ lattice show that the
deviation from the circle increases with the coupling strength, see
Fig.\ 2.
\begin{figure}[hbt]
\def\fpsangle{270}
\vspace{-14mm}
\epsfxsize=66mm
\fpsbox{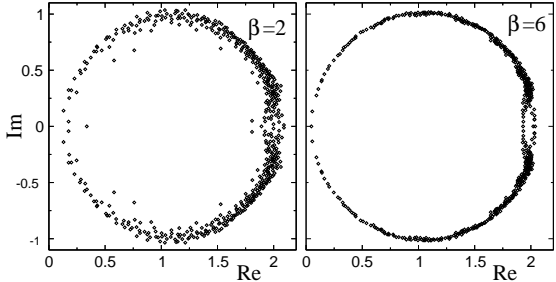}
\vspace{-32mm}
\caption{\it{Typical SO-HF spectra at strong and weak coupling,
approximating the GW unit circle.}}
\vspace{-8mm}
\end{figure}

As a scaling test we consider the dispersion
relations of the massless and the massive meson-type state -- which
we denote as $\pi$ and $\eta$ -- and we find again a strong improvement
of the SO-HF over the Wilson fermion, see Fig.\ 3. 
The SO-HF
reaches the same level as the classically perfect action \cite{LP},
although it only involves 6 different couplings per site (as opposed
to 123).
\begin{figure}[hbt]
\vspace{-8mm}
\def\fpsangle{0}
\epsfxsize=55mm
\fpsbox{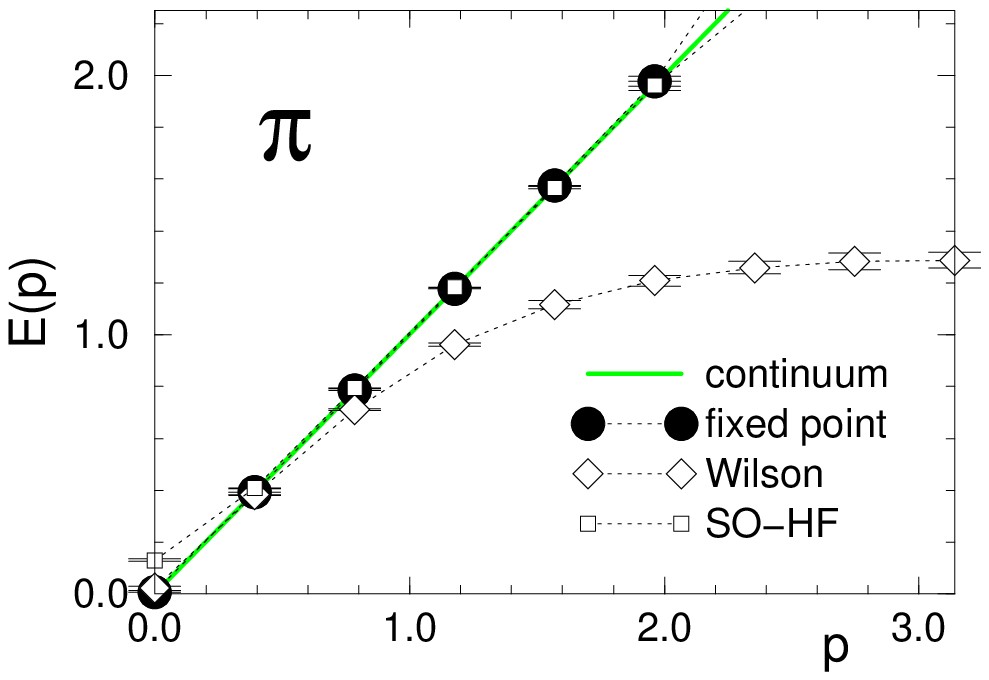}
\epsfxsize=55mm
\fpsbox{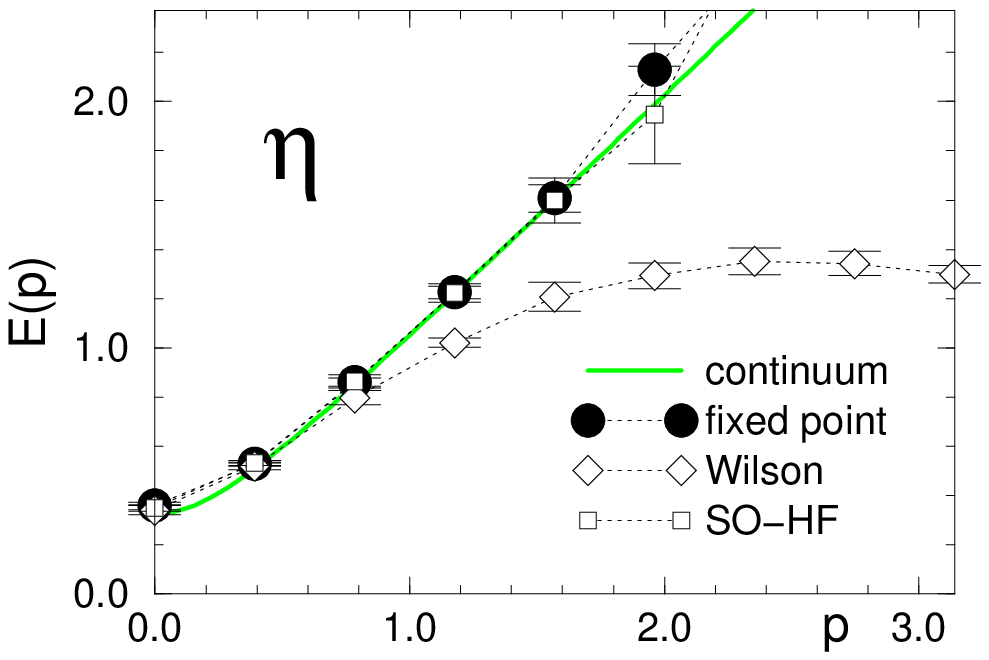}
\vspace{-12mm}
\caption{\it{The mesonic dispersions at $\beta=6$,
based on 5000 configurations. They are generated
quenched, but their evaluation does include the fermion determinant.}}
\vspace{-8mm}
\end{figure}

Hence the SO-HF has a remarkable quality with respect to chirality
and scaling, but the figures also show one unpleasant feature:
there is a strong additive mass renormalization, which leads to $m_{\pi}
\simeq 0.13$ (at $\beta =6$). 
There are various ways to move towards the chiral limit:
the standard method is to start from a negative bare mass, but one
can also achieve criticality solely by introducing fat links with
{\em negative} staple terms (which need to be tuned) \cite{BH}.
We now want to discuss yet another way, using the overlap
formula \cite{HN}.\\
\vspace{-3mm}

Our ansatz for $D$ obeys $D^{\dagger} = \gamma_{5} D \gamma_{5}$.
We now define $A = D-\mu$, thus the GWR (with $2R_{x,y} = \mu \ \delta_{x,y}$)
is equivalent to $A^{\dagger}A = \mu^{2}$. If we start from some $A_{0} =
D_{0}-\mu$, then the GWR does not hold in general. However, the operator can
be ``chirally corrected'' to $A = \mu A_{0} / \sqrt{ A_{0}^{\dagger} A_{0}}$.
Now $A$ does solve the GWR. H.\ Neuberger \cite{HN} suggested to 
insert the Wilson fermion $D_{0} = D_{W} = A_{W} +1$. 
We denote the fermion characterized
by $D_{Ne} = 1 + A_{W}/ \sqrt{ A_{W}^{\dagger} A_{W}}$ as Neuberger fermion.
At least in a smooth gauge background it is local (in the sense that its
couplings decay exponentially) \cite{HJL}.

The overlap-type of solution to the GWR can be generalized to a large
class by varying $D_{0}$ \cite{WB}. In particular, if $D_{0}$ represents
a GW fermion already (with $2 R_{x,y} = \mu \ \delta_{x,y}$), then
$D = D_{0}$. If we now insert a short-ranged, approximate (standard) GW fermion like
$D_{SO-HF}$, then it is altered only little by the overlap formula (with $\mu =1$), 
$D \approx D_{SO-HF}$. As a consequence, we can expect a high degree of locality
-- i.e.\ a fast exponential decay -- which is indeed confirmed in Fig.\ 4.
\begin{figure}[hbt]
\vspace{-8mm}
\def\fpsangle{0}
\epsfxsize=65mm
\fpsbox{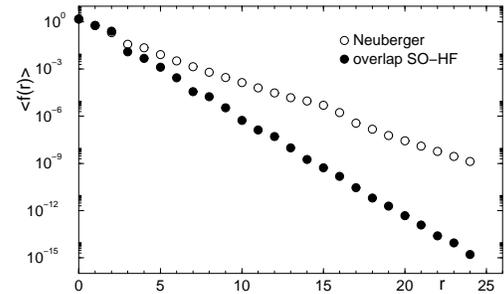}
\vspace{-12mm}
\caption{\it{The locality of the overlap SO-HF vs.\ Neuberger fermion,
measured by the maximal correlation $f(r)$
between two sites at distance $r$ (as suggested in Ref.\ [11])
on a $24 \times 24$ lattice.}}
\vspace{-9mm}
\end{figure}
Also the good scaling behavior of $D_{SO-HF}$ is essentially
preserved in the overlap SO-HF, see Figs.\ 1 and 5 for the fermionic and mesonic
dispersions. The same is true for the approximate rotational invariance \cite{BH}.
At the same time, we do have exact chiral properties now (hence $m_{\pi}=0$). 
If we look at the spectra of certain configurations before and after the use
of the overlap formula, the effect of the latter comes close to a radial
projection of the eigenvalues onto the GW circle.
\begin{figure}[hbt]
\vspace{-7mm}
\def\fpsangle{0}
\epsfxsize=55mm
\fpsbox{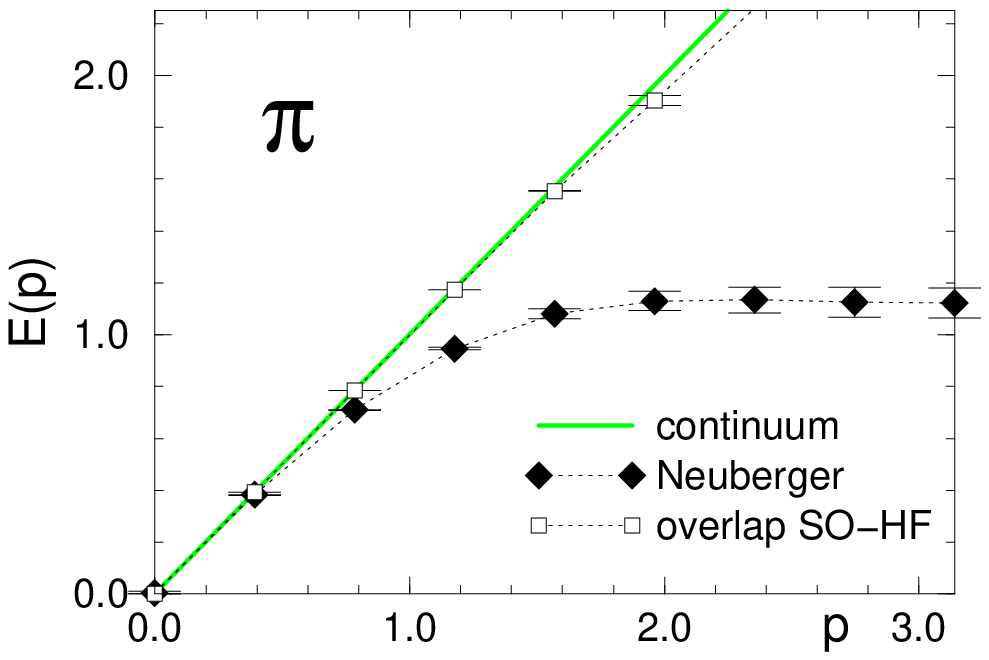}
\epsfxsize=55mm
\fpsbox{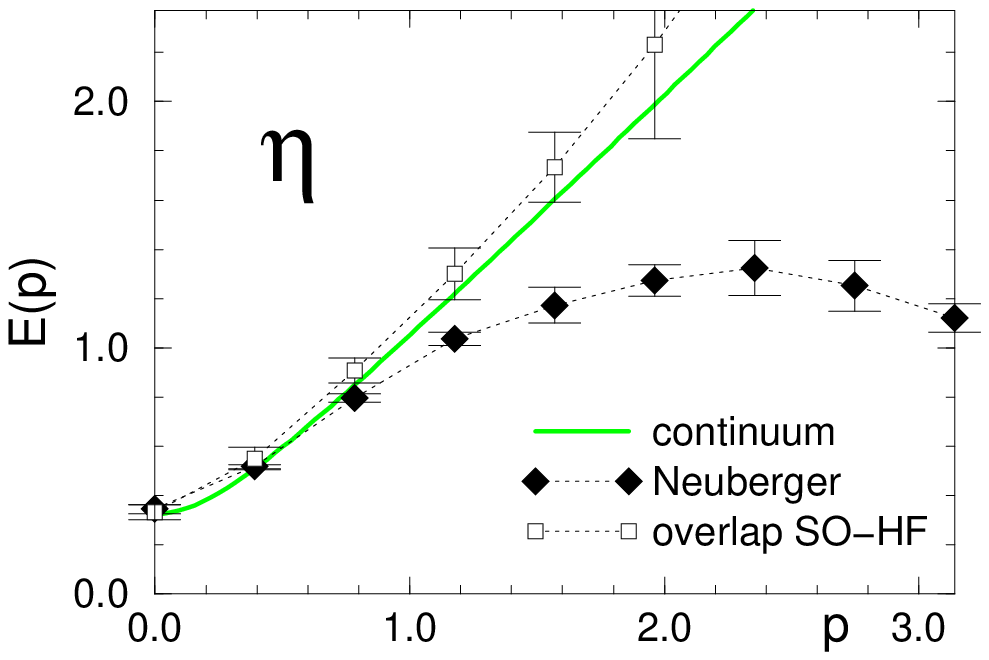}
\vspace{-12mm}
\caption{\it{The mesonic dispersions of different overlap fermions 
at $\beta=6$.}}
\vspace{-9mm}
\end{figure}

However, in QCD the use of the full improved overlap fermion might be
tedious due to the square root.
In view of $d=4$ we suggest to evaluate the square root in
\vspace*{-1.8mm}
\begin{displaymath}
D = \mu \ \Big[ 1 - A_{0}/\sqrt{ A_{0}^{\dagger} A_{0}} \Big]
\end{displaymath}
\vspace*{-1.8mm}
just perturbatively around $\mu$. For an approximate GW fermion
like $A_{0} = A_{SO-HF}$ this expansion converges rapidly, since
the operator $\varepsilon \doteq  A_{0}^{\dagger} A_{0} - \mu^{2}$
obeys $\Vert \varepsilon \Vert \ll 1$.
It fails to converge, however, for the Neuberger fermion.
If we perform the perturbative chiral correction, we obtain an operator
of the form $D_{p\chi c} = \mu - A_{0} Y$.
For the correction to $O(\varepsilon^{n})$ the operator $Y$ is given
by a polynomial in $A^{\dagger}_{0} A_{0} / \mu^{2}$ of order $n$.
The implementation then requires essentially $1+2n$ matrix-vector multiplications
(the matrix being $A$ resp.\ $A^{\dagger}$), hence the computational effort increases
{\em only} linearly in $n$.
It turns out that for moderate couplings already the leading orders are
efficient in doing most of the chiral projection, see Fig.\ 6.
\begin{figure}[hbt]
\def\fpsangle{0}
\epsfxsize=55mm
\fpsbox{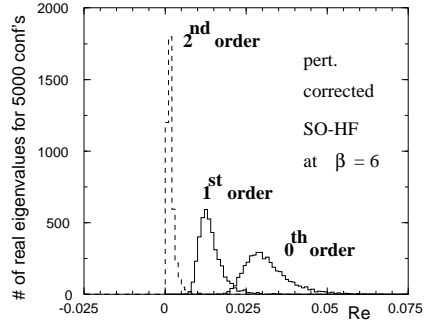}
\vspace{-11mm}
\caption{\it{
Histogram of the small real eigenvalues of $D_{p\chi c}$
-- based on $D_{0}= D_{SO-HF}$ -- 
showing that the mass renormalization vanishes quickly
under perturbative chiral correction.}}
\vspace{-9mm}
\end{figure}

Therefore, the perturbative chiral correction of a good HF combines
excellent scaling and chirality (and rotational invariance) 
as well as a high degree of locality with a
relatively modest computational overhead. 
This method is very promising for the extension to 4d, which
is currently in progress.

\end{document}